\documentclass[12pt,preprint]{aastex}

\shorttitle{Nonextensive probability distributions}
\shortauthors{Leubner and V\"{o}r\"{o}s}

\begin{document}


\title{A nonextensive entropy approach to solar wind intermittency}


\author{M. P. Leubner\altaffilmark{1} and Z. V\"{o}r\"{o}s }
\affil{Space Research Institute, Austrian Academy of Sciences,
    A-8042 Graz, Austria}
\email{manfred.leubner@oeaw.ac.at, zoltan.voeroes@oeaw.ac.at}


\altaffiltext{1}{also Institute for Theoretical Physics, University of Innsbruck,
A-6020 Innsbruck, Austria}


\begin{abstract}
The probability distributions (PDFs) of the differences of any
physical variable in the intermittent, turbulent interplanetary
medium are scale dependent. Strong non-Gaussianity of solar wind
fluctuations applies for short time-lag spacecraft observations,
corresponding to small-scale spatial separations, whereas for
large scales the differences turn into a Gaussian normal
distribution. These characteristics were hitherto described in
the context of the log-normal, the Castaing distribution or the shell model.
On the other hand, a possible explanation for nonlocality in turbulence is
offered within the context of nonextensive
entropy generalization by a recently introduced bi-kappa
distribution, generating through a convolution of a negative-kappa core
and positive-kappa halo pronounced non-Gaussian structures.
The PDFs of solar wind scalar
field differences are computed from WIND and ACE data for different
time lags and compared with the characteristics of the theoretical
bi-kappa functional, well representing the overall scale dependence
of the spatial solar wind intermittency. The observed PDF
characteristics for increased spatial scales are manifest in the
theoretical distribution functional by enhancing the only
tuning parameter $\kappa$, measuring the degree of nonextensivity
where the large-scale Gaussian is approached for $\kappa \rightarrow
\infty$. The nonextensive approach assures for experimental studies of
solar wind intermittency independence from influence of a priori
model assumptions. It is argued that the intermittency of the turbulent
fluctuations should be related physically to the nonextensive
character of the interplanetary medium counting for nonlocal
interactions via the entropy generalization. 
\end{abstract}




\section{Introduction}

Spatial intermittency in fully developed turbulence is an established feature of
astrophysical plasma fluctuations and is in particular manifest in the interplanetary
medium by in situ space probe observations. Hence, the analysis of probability
distribution functions (PDFs) or the computation of higher order moments are of
considerable interest to study multi-scale statistical properties of the solar
wind plasma and interplanetary magnetic field fluctuations. Classical statistical
theory provides the phase space distribution from ideal MHD invariants, introduced by
\citet{Matthaeus82} to characterize spectral properties of incompressible
solar wind fluctuations. The existence of multifractal structures in the
fluctuations of the magnetic field strength and the plasma parameters as
temperature, density and velocity of the interplanetary magnetofluid was
confirmed \citep{Burlaga91,Burlaga92}, followed by analyses of
non-Gaussian PDFs, related intermittency and the fractal scaling of the solar wind
MHD fluctuations by \citet{Marsch94,Marsch97}. The fractal/multifractal
nature of certain solar wind and magnetospheric key parameters, however, remains
to be a subject of recent investigations \citep{Hnat03}. Interplanetary observations
were used to establish the differences between fluid and MHD turbulence
\citep{Carbone95} and the Castaing distribution \citep{Castaing90,Castaing95,
Castaing96} was widely considered in favor of models of the
non-Gaussian cascade character of the intermittency in turbulent flows
\citep{Consolini98,Guowei98,Sorriso99,Schmitt01}. In addition to other solar
wind parameters as velocity and density the 
non-Gaussian and intermittent properties of the interplanetary magnetic
field fluctuations have shown to represent a new set of intermittency parameters
controlling the energy input rate into the magnetosphere \citep{Voros02a,Voros02b}.
Multi-scale intermittency and
anisotropy effects appear to be decisive factors in near-Earth magnetotail
dynamics as well \citep{Voros03,Voros04}. In the solar wind, based on WIND, ACE and
Voyager observations \citet{Burlaga02a,Burlaga02b} (and references therein)
provide a comprehensive study of multiscale statistical
properties of speed fluctuations at 1AU and 60 AU, respectively, verifying
that the PDFs at small scales have a shape that is characteristic
for intermittent turbulence with relatively large skewness and kurtosis
and approach a Gaussian for large scales (see also \citet{Sorriso99}). 

On observational grounds the evolution of energy spectra clearly demonstrates
the occurrence of nonlinear interactions in the solar wind fluctuations.
These interactions include turbulent cascades \citep{Tu95} as well as
nonlocal interactions between separated scales \citep{Chang99}.

The classical Boltzmann-Gibbs extensive thermo-statistics constitutes a
powerful tool when microscopic interactions and memory are short
ranged and the environment is a Euclidean space-time, a continuous and
differentiable manifold. However, in the present situation we are dealing
with astrophysical systems, generally subject to spatial or temporal
nonlocal interactions evolving in a non-euclidean, e.g.
multi-fractal space-time that makes their behavior nonextensive.
A generalization of the Boltzmann-Gibbs-Shannon entropy for a statistical
equilibrium was introduced by \citet{Tsallis88}, preserving the usual
properties of positivity, equiprobability and irreversibility, but
suitably extends the standard extensivity or additivity to nonextensivity.
With regard to plasma astrophysics we emphasize that the pseudo-additive
entropy concept generates power-law distributions where a parameter $q$
characterizes the fraction of the non-thermal particle components.
On the other hand, space physicists recognized four decades ago
from in situ satellite observations frequently the existence of high energy
tail velocity space distributions of electrons and ions in space plasmas.
The family of $\kappa-$distributions, a power law in particle speed, was
introduced on phenomenological grounds to model these suprathermal structures,
where $\kappa$ assumes only positive values. In continuation,
significant progress was provided by \citet{Treumann99a,Treumann99b} who developed
a kinetic theory demonstrating that power-law velocity space distributions are a
particular thermodynamic equilibrium state. Relating the parameters $q$ and $\kappa$ 
by an elementary transformation \citet{Leubner02} provided the missing link between
nonextensive distributions and $\kappa-$functions favored in space
plasma physics, leading to the required theoretical justification for
the use of $\kappa-$distributions from fundamental physics. Since the
parameter $\kappa$, a measure of the degree of nonextensivity of the
system, is not restricted to positive values in the nonextensive context,
the commonly observed core-halo twin character of the interplanetary
electron and ion velocity space distributions
was verified theoretically upon generalization to a bi-kappa
distribution, subject to a less pronounced core along with
extended tails as compared to a Maxwellian \citep{Leubner04a,Leubner04b}. 

Recently, the PDF of the Tsallis ensemble was linked to the analysis
of fully developed turbulence providing a relation between
nonextensivity and intermittency, which
manifests the multifractality in the distribution of eddies
\citep{Arimitsu00a,Arimitsu01}. It was shown that the value of the index $q$
of nonextensive statistics is related to the extremes of the multifractal
spectrum \citep{Lyra98} and that the multifractal spectrum corresponding to
the Tsallis statistics is determined self-consistently \citep{Arimitsu00b}.
Moreover, the context of generalized thermo-statistics provides
analytical formulas for PDFs of distance dependent velocity differences,
implying the cascade like structure of the turbulent dynamics and
an interpretation of the parameter $\kappa = 1/(q-1)$
\citep{Beck00}. Links between the nonextensive parameter $q$ and
the corresponding parameters of the log-normal, multifractal and
random-$\beta$ models were derived \citep{Shivamoggi03}.

We relate, in the following, nonlocality in turbulent flows to
nonextensive systems and demonstrate in the
context of entropy generalization the consistency of the theoretical
bi-kappa distribution \citep{Leubner04a} with observed, scale
dependent PDFs of characteristic key variables in the intermittent,
turbulent interplanetary medium, where $\kappa$ appears as the only
tuning parameter between the scales.

\section{Theory}

The nonextensive entropy generalization for the classical thermo-statistics
proposed by \citet{Tsallis88} takes the form

\begin{equation}
S_{q}=k_{B}\frac{1-{\sum }p_{i}^{q}}{q-1}
\label{1}
\end{equation}

where $p_{i}$ is the probability of the $i^{th}$ microstate, $k_{B}$ is
Boltzmann's constant and $q$ is a parameter quantifying the degree of
nonextensivity of the system, commonly referred to as the entropic index.
A crucial property of this entropy is the pseudo-additivity such that

\begin{equation}
S_{q}(A+B)=S_{q}(A)+S_{q}(B)+(1-q)S_{q}(A)S_{q}(B)
\label{2}
\end{equation}

for given subsystems $A$ and $B$ in the sense of factorizability of the
microstate probabilities. Hence, nonlocality or long-range
interactions are introduced by the multiplicative term accounting for
correlations between the subsystems. In order to link the Tsallis
$q-$statistics to the family of $\kappa-$distributions applied in
space and astrophysical plasma modeling we perform the transformation

\begin{equation} 
\frac{1}{1-q}=\kappa  
\label{3}
\end{equation}
  
to Eq. (\ref{1}) yielding the generalized entropy of the form \citep{Leubner02}

\begin{equation}
S_{\kappa }=\kappa k_{B}({\sum }p_{i}^{1-1/\kappa }-1)
\label{4}
\end{equation}

where $\kappa=\infty$ corresponds to $q=1$ and represents the extensive
limit of statistical independency. Consequently the interaction term
in Eq. (\ref{2}) cancels and the classical Boltzmann-Gibbs-Shannon
entropy is recovered as

\begin{equation}
S_{q}=-k_{B}{\sum p_{i}}\ln p_{i}
\label{5}
\end{equation}

Eq. (\ref{4}) applies to systems subject to nonlocal interactions
and systems evolving in a non-Euclidean and multifractal space-time. 
A further generalization of Eq. (\ref{2}) for complex systems composed of an
arbitrary number of mutually correlated systems is provided by
\citet{Milovanov00} where appropriate higher order terms appear in Eq. (\ref{2}).
Once the entropy is known the corresponding probability distributions are
available.

In Maxwell's derivation the velocity
components of the distribution $f(\mathbf{v})$ are uncorrelated where $\ln f$
can be expressed as a sum of the logarithms of the one dimensional distribution
functions. In nonextensive systems one needs to introduce correlations
between the components, which is
conveniently done by replacing the logarithm functions by a power law.
Upon transforming the "$q-exp$" and "$q-log$" functions introduced in
nonextensive statistics \citep{Silva98} into the $\kappa-$equivalents
$e_{\kappa}(f)=[1+f/\kappa]^\kappa$ and $ln_{\kappa} f=\kappa[f^{1/\kappa} -1]$ 
the generalized nonextensive solution for the one dimensional distribution
in $\kappa$ notation reads

\begin{equation}
f_{h}(v)=A_{h}\left[ 1+\frac{1}{\kappa }\frac{v^{2}}{v_{t}^{2}}\right]
^{-\kappa }
\label{6}
\end{equation}

Hence, the exponential probability function of the Maxwellian gas of an
uncorrelated ensemble of particles is replaced by the characteristics of
a power law. Note also that the distribution (\ref{6}) can be derived entirely
using general methods of statistical mechanics \citep{Milovanov00}.
Performing the normalization to the particle density $N$ yields
 
\begin{equation}
A_{h}=\frac{N}{\sqrt{\pi }v_{t}}\frac{1}{\sqrt{\kappa }}\frac{\Gamma %
\left[ \kappa \right] }{\Gamma \left[ \kappa -1/2\right] }
\label{7}
\end{equation}

and we assign to $f_{h}$ the notation 'halo' distribution.
$v_{t}=\sqrt{2k_{B}T/m}$ is the thermal speed
of a Maxwellian, where $T$ and $m$ are the temperature and mass,
respectively, of the species considered. The case $\kappa =\infty $
recovers the Maxwellian equilibrium distribution, see Fig. 1 and the
discussion below.

We note that contrary to Eq. (\ref{6}), derived from the nonextensive
entropy generalization Eq. (\ref{1}), the functional form of the $\kappa$-
distributions conventionally used in plasma physics is powered by $-(\kappa+1)$
instead of $-\kappa$ \citep{Leubner02,Leubner04a}. Consequently,
the nonextensive distribution (\ref{6}) provides more pronounced tails as
compared to the conventional $\kappa-$distribution for the same values of
$\kappa$. Furthermore, in all space plasma
applications the spectral index $\kappa$ was hitherto assumed to be restricted to
positive values, requiring in the derived one-dimensional nonextensive representation
$\frac{1}{2}<\kappa \leq \infty $ equivalent to $-1<q\leq 1$ (for $q<-1$ corresponding
to $\kappa < 1/2$ the distribution (\ref{6}) cannot be normalized). 

Since the nonextensive formalism is not restricted to $-1<q\leq 1$ we
generalize to values $q > 1$ providing the full range $-\infty \leq
\kappa \leq \infty $ for the spectral index $\kappa $. Hence, upon
incorporating the sign into the equation the one dimensional
distribution for negative definite $\kappa$ can be written in the form

\begin{equation}
f_{c}(v)=A_{c}\left[ 1-\frac{1}{\kappa }\frac{v^{2}}{v_{t}^{2}}\right]
^{\kappa }
\label{8}
\end{equation}

where the normalization constant reads

\begin{equation}
A_{c}=\frac{N}{\sqrt{\pi }v_{t}}\frac{1}{\sqrt{\kappa }}\frac{\Gamma %
\left[ \kappa +3/2\right] }{\Gamma \left[\kappa +1\right] }
\label{9}
\end{equation}

We assign the notation 'core' distribution to $f_{c}$, a function
subject to a thermal cutoff at the maximum allowed velocity
$v_{max}=\sqrt{\kappa}v_{t}$, which limits also any required integrations.
Since the sign is implemented into Eq. (\ref{8}), $\kappa$ has to be
understood as a positive quantity wherever it appears.

The nonextensive distribution-functionals (\ref{6}) and (\ref{8}), along
with the proper normalization from the entropy generalization in Eq. (\ref{1}),
provide access to nonlocal effects, where $\kappa$ measures the
degree of nonextensivity within the system.
Fig. 1 demonstrates schematically the non-thermal behavior of the
suprathermal halo component $f_{h}$ and the reduced core distribution $f_{c},$
subject to finite support in velocity space and exhibiting a thermal
cutoff at $v=v_{t}\sqrt{\kappa}$ where both functions approach the same
Maxwellian as $\kappa \longrightarrow \infty$.

With regard to a physical elucidation we refer to recent
developments indicating that the parameter $\kappa$ (or equivalently
$q$ via the transformation $1/(1-q)=\kappa$, as outlined above), as
degree of nonextensivity in the Tsallis context, finds at least three
more interpretations. In particular, \citet{Chavanis03}
illuminates the Tsallis q-entropy approach in view of
incomplete mixing in complex systems, clarifying that it is related to
a particular Boltzmann H-function, where the parameter $q$ measures
the degree of mixing. Furthermore, the PDFs of the Tsallis
ensemble are linked to the analysis of fully developed turbulence
\citep{Arimitsu00a,Arimitsu01}, where the nonextensive parameter $q$
is in particular related to nonlocality in the system and
measures the scale dependence of the PDFs
of the dynamical physical variables considered. Finally, the parameter $q$
may find also a possible alternative physical interpretation in terms of the
heat capacity of a medium. A system with $\kappa < 0$ was suggested to
represent a heat bath with finite positive heat capacity and vice versa,
for $\kappa > 0$ the heat bath has negative finite heat capacity. For a
discussion of this property with regard to entropy generalization see
\citet{Almeida01}. This outline indicates that the conceptual background
for the description of nonextensive systems requires the development
of unifying concepts.  

It should be emphasized that, besides a brief comment \citep{Leubner02},
the extension to negative $\kappa-$values as formulated in Eq. (\ref{8})
was just recently recognized as being physically relevant
\citep{Leubner04a,Leubner04b} since research focused hitherto on the suprathermal
particle populations, in particular, well represented by a
halo distribution. Let us point out that the thermal cutoff of the functional
Eq. (\ref{8}) provides apparently a clear definition of a core population,
whereas a 'core' of the function $f_{h}$ is not definable and just in some
arbitrary way related to the central part of a Maxwellian. Besides a clearly defined
core-halo transition any unique and physically relevant nonextensive PDF must
be subject to three further conditions. We require that (a) the distribution
approaches one and the same Maxwellian as $\kappa \longrightarrow \infty,$
(b) a unique, global distribution must be definable by one single density and temperature 
and (c) upon variation of the coupling parameter $\kappa$ particle conservation
and adiabatic evolution should be considered, such that a redistribution in a box
(a source free environment) can be performed. Subject to these constraints the
appropriate mathematical functional, representing observed core-halo structures in
nonextensive astrophysical environments, is available from the elementary combination 
$F_{ch}=B_{ch}(f_{h}+f_{c})$. In this context the full velocity space distribution
compatible with nonextensive entropy generalization and obeying the above
constraints reads

\begin{equation}
F_{ch}(v;\kappa)=\frac{N}{\pi^{1/2}v_{t}}G(\kappa)
\left\{\left[1+ \frac{1}{\kappa} \frac{v^{2}}{v_{t}^{2}}\right]^{-\kappa}  
+\left[ 1- \frac{1}{\kappa} \frac{v^{2}}{v_{t}^{2}}\right] ^{\kappa }\right\}
\label{13}
\end{equation}

to which we assign the notation 'bi-kappa distribution' and where
$B_{ch}=N/(\pi^{1/2}v_{t})G(\kappa)$ and the function $G(\kappa)$
is defined by

\begin{equation}
G(\kappa)=\left[ \frac{\kappa^{1/2}\Gamma (\kappa-1/2)}{\Gamma (\kappa)}+
\frac{\kappa^{1/2}\Gamma(\kappa+1)}{\Gamma (\kappa +3/2)}\right]^{-1}
\label{14}
\end{equation}

from the normalization. The function $G(\kappa)$ is subject to a particular
weak $\kappa-$dependence where $G(\kappa) \simeq 1/2$, see \citet{Leubner04a}
for a graphical illustration and discussion. Hence, the resulting factor $1/2$ in the
normalization constant $B_{ch}=N/(2\pi^{1/2}v_{t})$ reflects consistently the superposition
of the two counter-organizing contributions in Eq. (\ref{13}). Upon re-distribution of
particles as a consequence of nonlocal interactions, where $\kappa$ act as measure of
nonlocality in the system, a fraction of core-particles is removed and available
to generate a pronounced non-Maxwellian tail population counting for zero net
particle budget. Hence, the specific nonextensive velocity space
distribution (\ref{13}) provides a unique mathematical core-halo representation
subject to the non-trivial property of particle conservation in the sense that the
normalization is independent of the parameter $\kappa$ \citep{Leubner04a,Leubner04b}.
This specific feature allows also to transform the one-dimensional
bi-kappa velocity distribution of Maxwell's particle context into the mathematical
form of a Gaussian of constant variance, a PDF applicable to any characteristic
physical variable suitable to analyze statistical properties of astrophysical systems,
see below. Contrary to the conventional $\kappa-$distribution models
constraint by a positive $\kappa-$value, the functional form (\ref{13})
permits in a closed system a re-distribution of energy from a Maxwellian into highly
non-thermal core-halo structures where no support from the environment is required. 
This particular physical feature is only available if $\kappa$ assumes the same
value for both, core and halo fraction and is manifest simultaneously in a 
core-width reduction and distinct tail formation as $\kappa$ decreases. For
$\kappa = \infty$ the function $G(\kappa) = 1/2$ and
the power laws in the brackets of the right hand side of Eq. (\ref{13})
turn each into the same Maxwellian exponential.
 
Finally, besides particle conservation also adiabatic behavior is known as typical
characteristic from a variety of astrophysical plasmas studies
\citep{Baumjohann89,Totten95}. The effect of long-range interactions
on the entropy dependence $S(\kappa)$ in a system obeying a bi-kappa PDF
(\ref{13}) can be studied after a decomposition into the core and halo
fractional entropy contributions \citep{Leubner04a}. Nonlocality in the system
generates, as consequence of the pronounced tails, an
entropy increase with regard to the halo PDF $f_{h}$ that is
accompanied by a counteracting partner, the core entropy contribution
originating from the cutoff PDF $f_{c}$, a higher organized state of
reduced entropy as compared to the Maxwellian equilibrium entropy. 
In particular, the entropy of the full bi-kappa distribution Eq. (\ref{13}) 
follows the Boltzmann-Gibbs-Shannon extensive behavior down to $\kappa \sim 10$
turning after a transition into the nonextensive profile of $f_{h}$ alone.
Since $dS \geq 0$ as $\kappa$ decreases from $\kappa = \infty$ to low
values any generation of highly non-Maxwellian PDF structures due to
nonlocality in multifractal systems obeys consistently the second law.

Intermittency in astrophysical plasma turbulence can be analyzed 
by comparing observed PDFs of fluctuations of the velocity field magnitude in
the solar wind with the global theoretical core-halo representation (\ref{13}),
accounting for nonlocality through entropy generalization. 
On the other hand, on small scales also the PDFs of interplanetary density and magnetic
field fluctuations are observed to obey the same leptokurtic core 
non-Gaussianity along with pronounced tails or halos, requiring a generalization
of Eq. (\ref{13}).

We introduce a unique, global PDF provided in  the context of nonextensive
thermo-statistics for any physical
variable, where nonlocality is associated to pseudo-additive
entropy generalization. Upon normalizing the one-dimensional
bi-kappa particle distribution (\ref{13}) to unity and assigning a 
unique distribution variance $\sigma$ to the thermal spread $v_{t}$, the
'Maxwellian form' of the bi-kappa distribution (\ref{13}) transforms
with $G(\kappa) = 1/2$ to a 'Gaussian form' of a global bi-kappa PDF as

\begin{equation}
P_{ch}(\delta X;\kappa)=\frac{1}{2\pi^{1/2} \sigma}
\left\{\left[1+ \frac{1}{\kappa} \frac{\delta X^{2}}{\sigma^{2}}\right]^{-\kappa}  
+\left[ 1- \frac{1}{\kappa} \frac{\delta X^{2}}{\sigma^{2}}\right] ^{\kappa }\right\}
\label{16}
\end{equation}

Subject to a single constant variance $\sigma$ this one-parameter
PDF is applicable to the differences of the fluctuations 
$\delta X(\tau)=X(t+\tau)-X(t)$ of any physical variable $X$ in the
astrophysical system considered, where $X(t)$ denotes any characteristic
solar wind variable at time $t$ and $\tau$ is the time lag. 
Here $\kappa$ assumes a clear physical
interpretation defining the degree of nonextensivity or nonlocality
in the system, thus being a measure of long-range interactions. As
$\kappa \longrightarrow \infty$ the bi-kappa distribution $P_{ch}(\delta X;\kappa)$
approaches a single Gaussian. In the following we demonstrate the relevance
of the nonextensive context by analyzing the observed scale
dependence of the PDFs of the differences of velocity, density and
magnetic field variables in the intermittent, turbulent interplanetary
medium based on the theoretical global bi-kappa PDF (\ref{16}).  
 
\section{Data Evaluation}

For a proper estimation of empirical PDFs of interplanetary plasma and
magnetic field data long sets are needed. Due to non-stationarity in
driving dissipation conditions, however, the data sets have to be chosen
carefully. For the present approach solar wind velocity data
are analyzed with a time resolution of 92 s  and the
interplanetary magnetic field strength with a time resolution
of 60 s, both available from the WIND spacecraft.
WIND has a complicated trajectory that also crosses the magnetosphere from
time to time. Our analysis is based on the entire year 1997 data set when WIND
was in the GSE positions: 150-230 $R_{E}$. This period is compared with a shorter
time period from November 18 to December 10, 1998 when magnetic 
field measurements with time resolution of 16 s were available from the ACE
spacecraft. The ACE satellite is continuously
monitoring the solar wind at the L1 point. Data gaps were interpolated in all cases and
the resulting PDFs are fitted by the theoretical nonextensive distributions.
Moreover, we provide a comparison of the WIND magnetic field analysis with
the corresponding PDFs obtained from ACE magnetic field measurements.
In particular, from the 1997 WIND velocity and magnetic field data 5 day periods
of low speed subintervals were selected such that during each subinterval the
bulk velocity was less than 450 km/s. 

No velocity observations were available from ACE during the interval of January
10-29, 1998. However, at this time WIND was upstream, close to the ACE position.
WIND measurements provide clear evidence that between November 18 and December 10, 1998, the
bulk speed was less than 450 km/s in about 92 \% of this period. This interval was treated
in one block as a whole since also the magnetic field intermittency was low,
where any increase of the latter can be associated with high speed streams and
increased geomagnetic response \citep{Voros02a}.

For each data set the plasma and magentic field increments were calculated at
a given time lag $\tau$, equivalent to a particular spatial scale by
$\delta X(\tau)=X(t+\tau)-X(t)$ where each variable within each subinterval
is normalized to the standard deviation. For each 5 day long realization
of the low speed flow the empirical probability distribution function (histogram) was
then computed. $\delta X(\tau)$ is binned into $N$ equal spaced boxes and the number
of elements in each box was computed where the robustness of the histograms against
$N$ is tested. $\delta X(\tau)$ represents characteristic fluctuations at the time
scale $\tau$ or equivalently across eddies of size $l \sim V*\tau$. In the case of
single spacecraft measurements a similarity in statistical properties of temporal
or spatial fluctuations is supposed. This is true in particular for the solar wind
\citep{Marsch97} where the spatial fluctuations on a scale $l$ pass over the
spacecraft faster than they typically fluctuate in time. Hence, by changing $\tau$
it is possible to analyze the statistical features of fluctuations in different
time scales, which roughly correspond to those statistical characteristics
across turbulent eddies of size $l \sim V*\tau$. 

At large scales, e.g. $\tau= 24$ hrs. and $l\sim 5000 R_{E}$ (supposing $V=350$ km/s)
the PDF's are close to the Gaussian distribution, because the correlation
between the separated points vanishes, whereas at small scales, e.g. $\tau= 0.2$ hrs.
and $l\sim 40 R_E$, the PDF's clearly have non-Gaussian shapes, see section 4. These
values were also estimated by \citet{Sorriso99}.

\section{Results and Discussion}

Core-halo structures are a persistent feature of a variety of different
astrophysical systems. In particular, a manifestation of those formations
is found in the thermo-statistical properties
of interplanetary plasmas (e.g. \citet{Marsch82,Pierrard99,Maksimovic00}),
in gravitational equilibrium states of stellar system \citep{Chavanis98} or in the
distribution of dark matter in galaxies and beyond, see e.g. \citet{Burkert95,
Salucci00}. We focus in this study on astrophysical plasma
turbulence, specifically on PDFs of fluctuations in the intermittent solar
wind turbulence, safely available from in situ satellite observations.

The Wind and ACE solar wind data analysis unambiguously manifests that 
the PDFs of large scale velocity, density and magnetic field fluctuation differences
are well represented by a Gaussian, turning simultaneously into leptokurtic
peaked distributions of strong non-Gaussianity in the center along with a 
pronounced tail structure on small scales. In particular, the PDFs of large-scale
magnetic field fluctuations, not related to the increment field are known to be
subject to relatively small deviations from the Gaussian statistics
and are well fitted by the Castaing distribution, a convolution of
Gaussians with variances distributed according to a log-normal distribution
\citep{Castaing90,Padhye01}.
Independent of the physical situation considered, the Castaing distribution
provides a multi-parameter description of observed PDFs, plausible in this case,
since the large scale fluctuations of the interplanetary magnetic field are
generated by a variety of discrete coronal sources. If individual coronal sources
evoke Gaussian distributed magnetic fields, the net magnetic fluctuations can be
modeled by their superposition with a spread of the corresponding variances.
Here the Castaing model is not related to any energy cascading processes.

Contrary, small-scale fluctuations of the increment field are associated with local
intermittent flows where fluctuations are concentrated in limited space volumes. Consequently,
the PDFs are scale dependent and intermittency generates long-tailed distributions.
It is customary to use $n$-th order absolute powers of the plasma variables
and magnetic field increments ($n$-th order structure functions, \citet{Pagel01,Marsch97})
allowing to investigate the multi-scale scaling features of fluctuations
subject to long-tailed PDFs. Direct studies of observed PDFs of the increment fields
$\delta X(\tau)=X(t+\tau)-X(t)$ for any characteristic solar wind  
variable at time $t$ and time lag $\tau$ revealed departures from a Gaussian
distribution over multiple scales \citep{Sorriso99} and an increase of intermittency
towards small scales \citep{Marsch94}. The PDFs are also found to be leptokurtic 
which indicates the turbulent character of the underlying fluctuations. 
\citet{Sorriso99} have shown that the non-Gaussian behavior of small-scale velocity
and magnetic field fluctuations in the solar wind can also be described well by a
Castaing distribution, where the individual sources of Gaussian fluctuations appear as
a fragmentation process through scales in turbulent cascades.

A more general fragmentation process based on the extreme deviation theory
was proposed by \citet{Frisch97}. In this model the increment fields are due to
the product of fragments derived from structures at larger scales and the PDF
is a stretched exponential function $P_{se}(\delta X) \sim exp[-\beta |\delta X|^{\mu}]$,
where $\beta$ and $\mu$ are the tunable parameters, governing the
fragmentation process. \citet{Sorriso01} have shown
that both, the Castaing and the stretched exponential model can describe
intermittency in the solar wind equally well. Fig. 2 provides a comparison of
the stretched exponential model, the widely used log-normal model \citep{Frisch96}
$P_{ln}(\delta X) \sim 1/(\sigma \delta X)exp[-ln(\delta X)^{2}/\sigma^{2}]$, 
where $\sigma$ is the variance of the distribution and $m$ is the scale parameter,
both compared to a bi-kappa fit to ACE magnetic field
amplitude data between November 18 and December 10, 1998.
The dimensionless $\tau=14$ is multiplied
by the time resolution of 16 sec. to generate an effective time-lag and
the log-normal model restricts to positive values of the differences. All three
models provide excellent fits to the data indicated by the error bars
for large differences, whereas the stretched exponential model slightly 
deviates from experimental curve at small $\delta B$.  

In Fig. 3 we focus on the bi-kappa model showing for the same date set
that the scale dependence of the PDF in the solar wind can be represented
accurately via the only tuning parameter $\kappa$, which measures the
degree of nonextensivity or coupling within the astrophysical plasma
environment. The nonextensive context, generates a precise representation
for the observed PDFs characterizing the intermittency of the small
scale fluctuations. For comparison also a Gaussian and the conventional
$\kappa-$function \citep{Leubner02}, subject to the same $\kappa-$value
and not able to reproduce the structure of the PDF of small scale fluctuations,
are provided. Here and in the following figures the dotted lines
indicate the standard deviation.

Based on WIND velocity field magnitude data Fig. 4 presents an analysis of the
scale dependence of interplanetary PDFs of the velocity field magnitude
indicating in three plots from left to right the decreasing
intermittency with increasing spatial separation scale or time-lags,
where now the observational uncertainty is indicated by the standard
deviation (dotted lines). Constraint by a time resolution of 92 sec.
the two point time separation is $\tau \times 92 sec$ where $\tau$
assumes the values 10, 70 and 900 from left to right. The solid
lines represent best fits to the observed PDFs with nonextensive
distributions where the corresponding $\kappa$ is determined as
$\kappa = 2, 3.5$ and $\infty$. The strong non-Gaussian character of
the leptokurtic PDFs (left panel) exhibiting pronounced tails associated with 
solar wind turbulence and intermittency in small scale fluctuations
finds again an accurate analytical fit and hence a physical
background in the nonextensive representation, accounting for nonlocality
in turbulence. The non-Gaussian structure
is somewhat softened for enhanced $\tau = 70$ (central panel) but
again precisely modeled within the pseudo-additive entropy context
turning into the Gaussian shape of large scale fluctuations, independent
of the increment field.

Fig. 5 and Fig. 6 provide the corresponding nonextensive analysis of the
scale dependence of the density and magnetic field fluctuations obtained
from WIND data. Evidently, the scale dependent characteristics of the
observed PDFs of the increment fields $\delta X(\tau)=X(t+\tau)-X(t)$
for all solar wind variables evolve simultaneously on small scales
approaching independency of the increment field in the large scale
Gaussian. With separation scales of $\tau = 10, 70, 900$ the
corresponding evolution of the PDFs of observed density and
magnetic field fluctuations are best represented by the same values
of $\kappa = 2, 3.5, \infty$ counting for nonlocal interactions
in the nonextensive theoretical approach as for the velocity field
magnitude. 

Finally, Fig. 7 demonstrates from ACE magnetic field observations that
the results of the structure analysis regarding the scale dependence
of the increment field PDFs of physical variables in the interplanetary
medium performed by WIND data is not restricted to the observations
of this particular spacecraft. Highly accurately, the overall scale
dependence appears as a general characteristic of quiet astrophysical
plasma environments. In particular, the scale dependent PDF evolution
of magnetic field fluctuations is subject to a two point separation
scale of $\tau = 100, 2000, 10000$ with a time resolution of 16 sec.
Best fits of the bi-kappa distribution are obtained for
$\kappa = 1.8, 3, \infty$, measuring the degree of nonextensivity,
or coupling through nonlocal interactions, respectively.             
     
Figs. 4, 5 and 6 demonstrate clearly via the PDFs of the density and
the velocity and magnetic field magnitudes that the scale dependence
of the fluctuations of different characteristic interplanetary
variables exhibits the same behavior or is at least similar. The
analysis is based on slow speed solar wind PDFs indicating a
universal scaling dependency between N,v and B intermittency
within the experimental uncertainties. 
This strong correlation implies that the scale dependencies
of all physical variables are coupled where the solar wind Alfv\'{e}nic 
fluctuations provide a physical basis of the velocity and magnetic
field correlations. Furthermore, the proportionality between
density fluctuations and the magnetic field and velocity fluctuations is
already maintained in the solar wind by the presence of weak spatial
gradients \citep{Spangler03}.

$\kappa-$distributions reproduce the Maxwell-Boltzmann distribution for
$\kappa \rightarrow \infty$, a situation identifying $\kappa$ as an ordering
parameter that acounts for correlations within the system. Quasistationary,
highly correlated turbulent conditions characterized by kappa distributions
represent stationary states far from equilibrium where a generalization of the
Boltzmann-Shannon entropy, as measure of the level of organization or
intermittency, applies \citep{Goldstein04,Treumann04}. Physically this can be understood
considering a system at a certain nonlinear stage where turbulence may reach a
state of high energy level that is balanced by turbulent dissipation. In this
environment equilibrium statistics can be extended to dissipative systems, approaching
a stationary state beyond thermal equilibrium \citep{Gotoh02}. However, no final
theory exists for the problem of nonequilibrium statistics. Since turbulence
is driven in the solar wind by velocity shears we choose for the data analysis
intervals of low speed solar wind with limits in velocity space, where the
driving and dissipation conditions do not change significantly, maintaining
therefore the dynamical equilibrium condition.

MHD turbulence in the solar wind is generally in a state in which propagating
waves and non-propagating fluctuations coexist \citep{Shebalin83}. Moreover, typical
processes include fluctuation induced multiscale interactions of coherent structures
for which a local interaction in Fourier space is not required and interactions
of opposite traveling waves are followed by energy cascades in Fourier space.
Physically, at least a part of the nonlocal forces and turbulent intermittency
results from a source exhibiting nonlocal interactions in the Fourier space.
We proposed a model for describing those nonlocal interactions on the basis of
entropy generalization. However, alternative models accounting for nonlocality,
responsible for intermittency in general, are available e.g.
\citep{Tsinober90,Chang99,Laval01,Falkovich01} demonstrating
that the origin of intermittency due to nonlocal interactions,
in particular in astrophysical plasmas as the solar wind, remains open. 
   
Summarizing, a bi-kappa distribution family turns out theoretically
as consequence of the entropy generalization in nonextensive
thermo-statistics and was successfully applied in velocity space
to interplanetary core-halo electron distributions
as well as double humped proton structures. Upon transforming
to the mathematical 'Gaussian form' the particular global,
on-parameter bi-kappa PDF of constant variance provides also access to
the pronounced core-halo distributions observed in astrophysical
plasma turbulence. A redistribution of a Gaussian on large scales can
be performed physically in a closed environment turning into
highly non-Gaussian leptokurtic and long-tailed structures. Pseudo-additive
entropy generalization provides the required physical interpretation of
the only tuning parameter $\kappa$ in terms of the degree of
nonextensivity of the system as a measure of nonlocality or coupling.
The scale dependence of the PDFs of the differences
of characteristic physical variables in the intermittent turbulent
interplanetary medium as density, velocity and magnetic field magnitude
is accurately represented by the nonextensive bi-kappa
functional. It is argued that the leptokurtic, long-tailed non-Gaussian
core-halo PDFs characterizing the intermittency of the turbulent fluctuations
can be related to the nonextensive character of the interplanetary
medium via the entropy generalization.

\acknowledgments

Acknowledgments: 
M. Leubner acknowledges the hospitality of the Austrian Academy of Sciences
at the Space Research Institute in Graz.
The authors are grateful to N. Ness (Bartol Res. Inst.) 
for providing ACE data and to R. Lepping and K.W. Ogilvie (NASA-GSFC) for
providing WIND data and M. Volwerk for reading the manuscript.

\clearpage


\begin{figure}
\plottwo{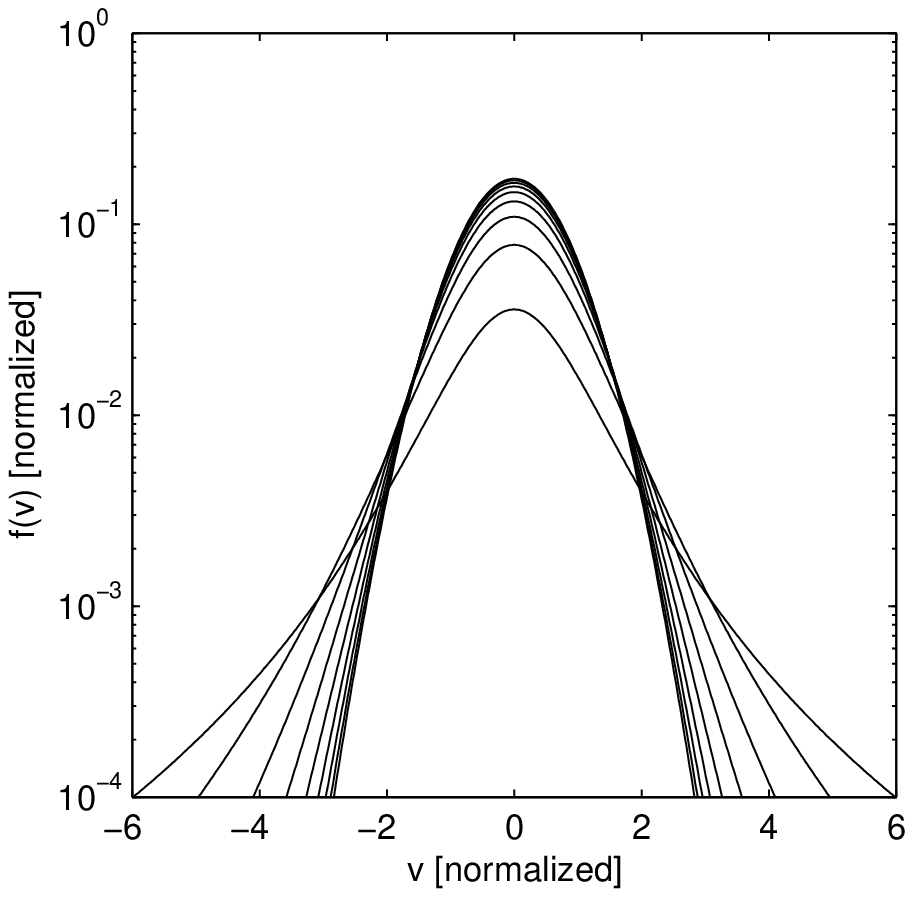}{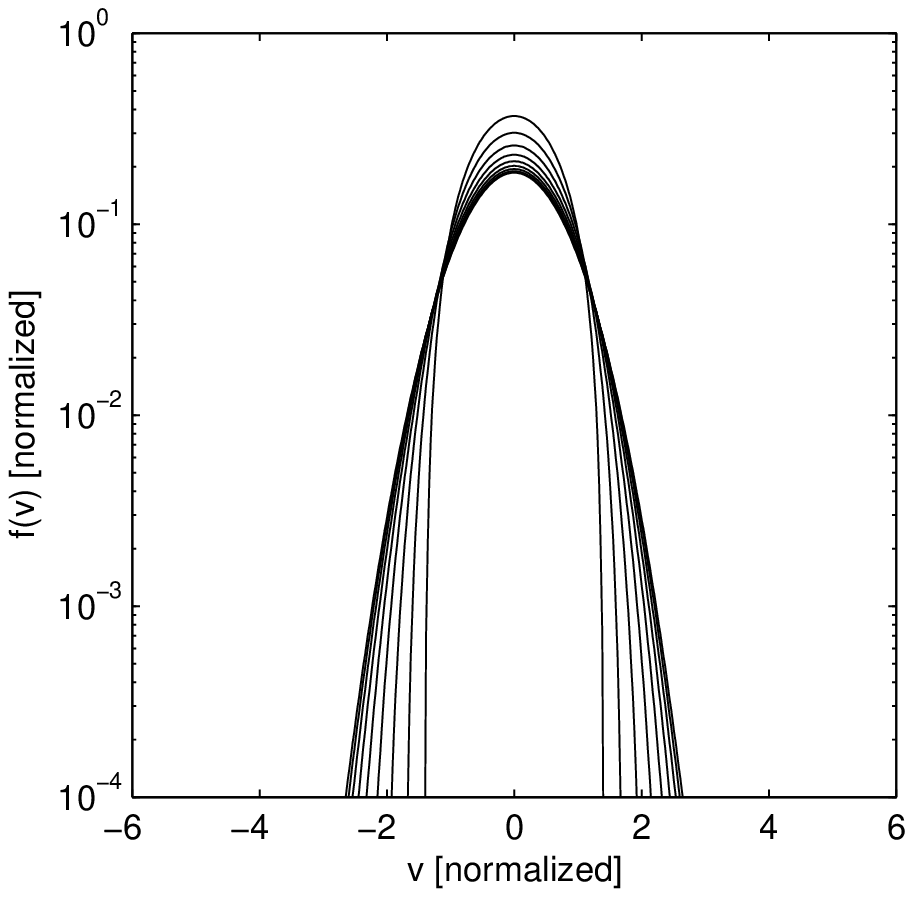}
\caption{A schematic decomposition of the characteristics of the nonextensive
bi-kappa distribution family Eq. (\ref{13}):
with $\kappa=3$ the outermost curve (left panel) and innermost curve (right panel)
correspond to the halo $f_{h}$ and core $f_{c}$ distribution fraction.
For increasing $\kappa-$values both sets of curves
merge at the same Maxwellian limit indicated as bold line,
$f_{h}$ from outside and $f_{c}$ from inside.
\label{fig1}}
\end{figure}

\clearpage 

\begin{figure}
\plotone{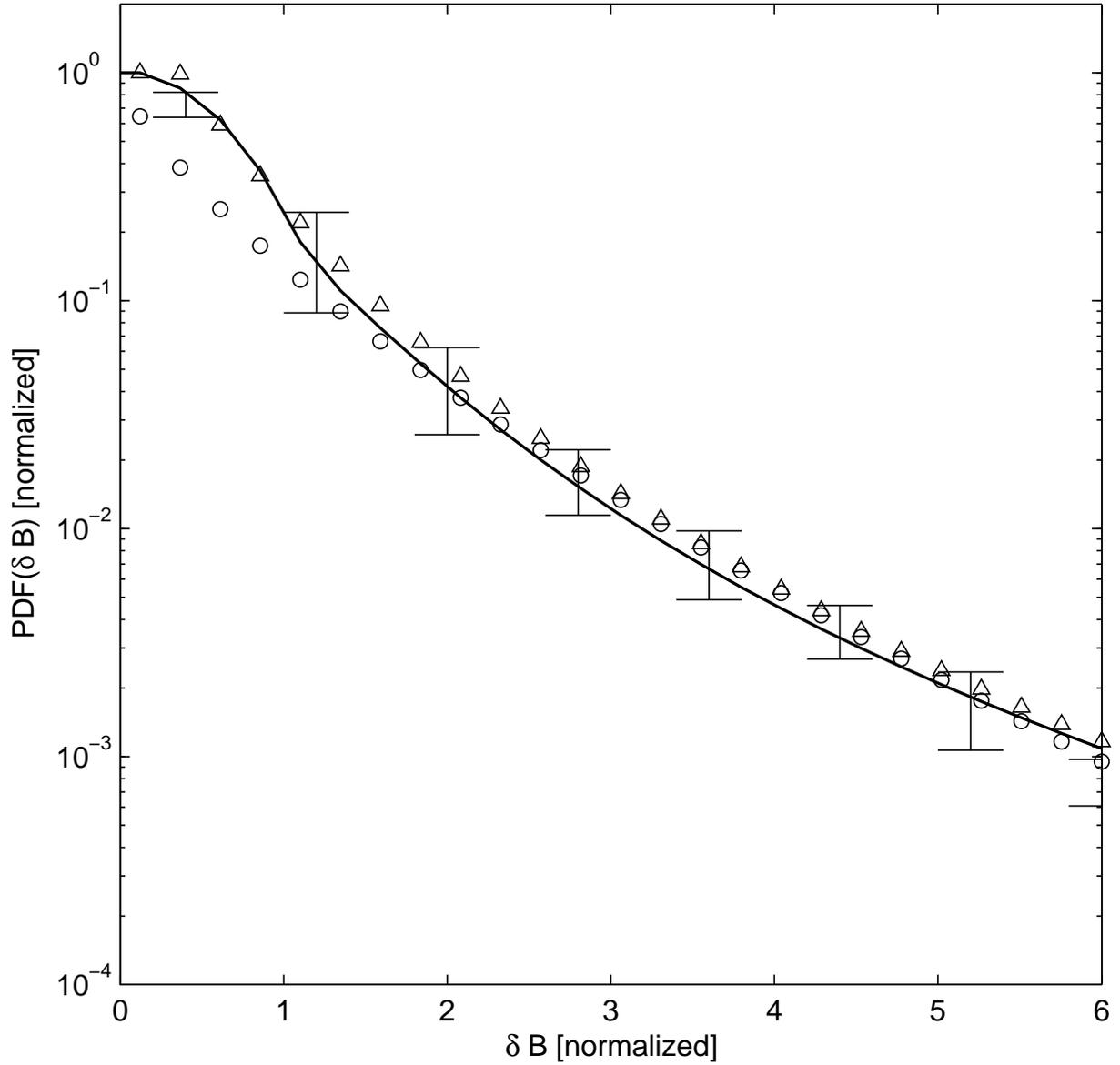}
\caption{A comparison of a stretched exponential (circles, $\beta=1.95, \mu=0.71$)
and a log-normal (triangles, $\sigma=0.9, m=0.75$) distribution with the
nonextensive bi-kappa function (solid line, $\kappa=1.8$)
and ACE magnetic field fluctuation differences (error bars). 
\label{fig2}}
\end{figure}

\clearpage 

\begin{figure}
\plotone{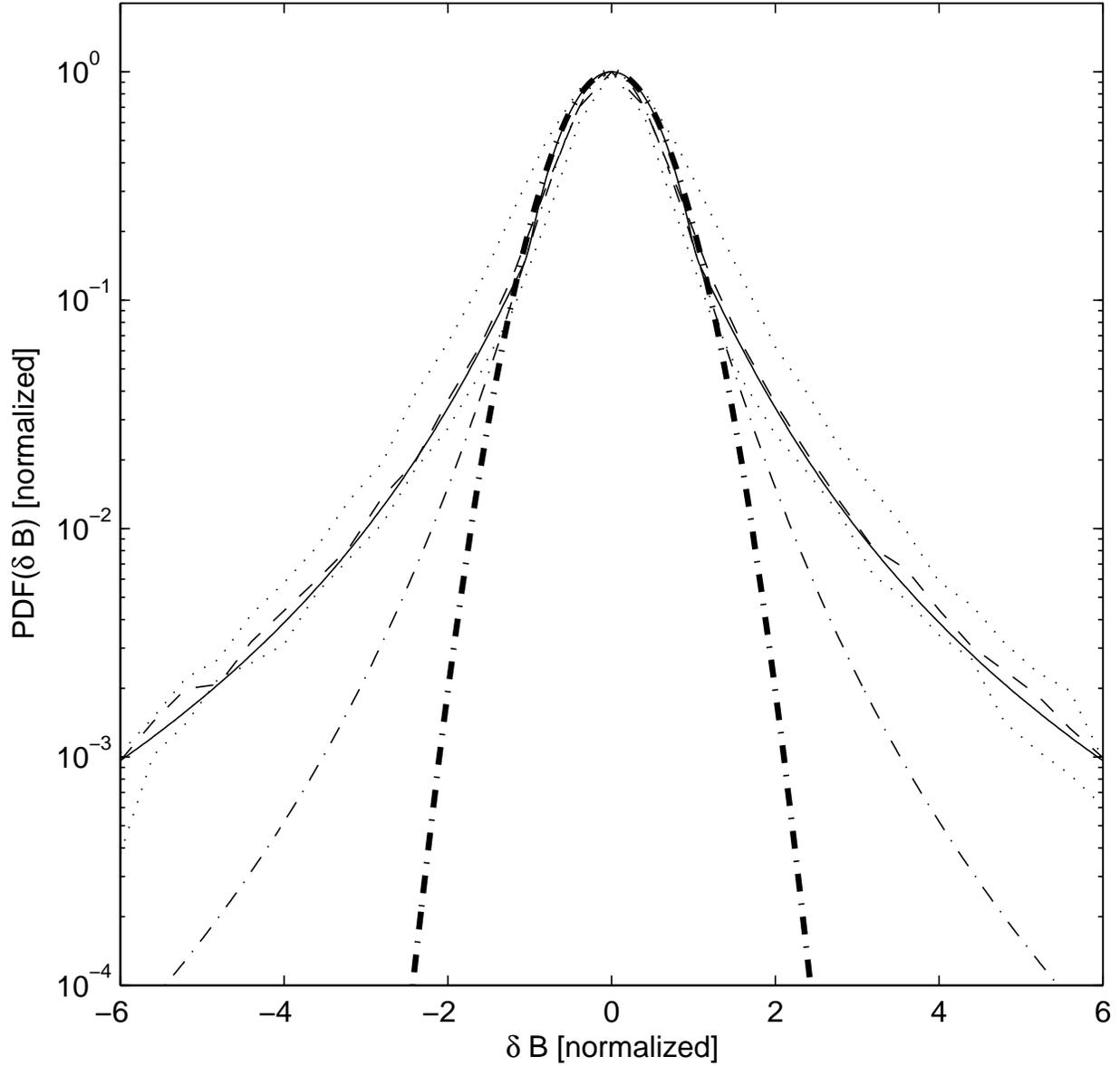}
\caption{A nonextensive bi-kappa fit (solid line) with $\kappa=1.8$
of an observed PDF (dashed line) obtained from ACE magnetic field amplitude
data. A Gaussian (thick dashed-dotted line) and a conventional
$\kappa-$distribution (thin dashed-dotted line) are 
provided for comparison. The dotted lines indicate the standard deviation.
\label{fig3}}
\end{figure}

\clearpage 

\begin{figure}
\plotone{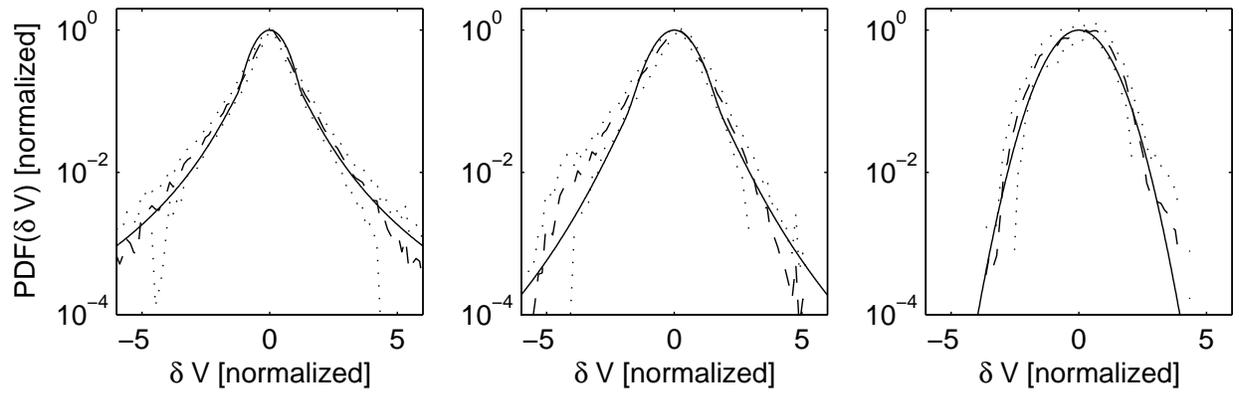}
\caption{Left panel: The PDF of the increments of observed WIND velocity
field magnitude fluctuations for $\tau = 10$ and a resolution of 92 sec.
as compared to the bi-kappa function with $\kappa=2$. Based on the same
data the central panel provides the characteristics for increase $\tau = 70$
where $\kappa$ assumes a value of 3.5 for the best representation. The PDF
of large scale velocity fluctuations are well modeled by a Gaussian with
$\kappa = \infty$, right panel.  
\label{fig4}}
\end{figure}

\clearpage 

\begin{figure}
\plotone{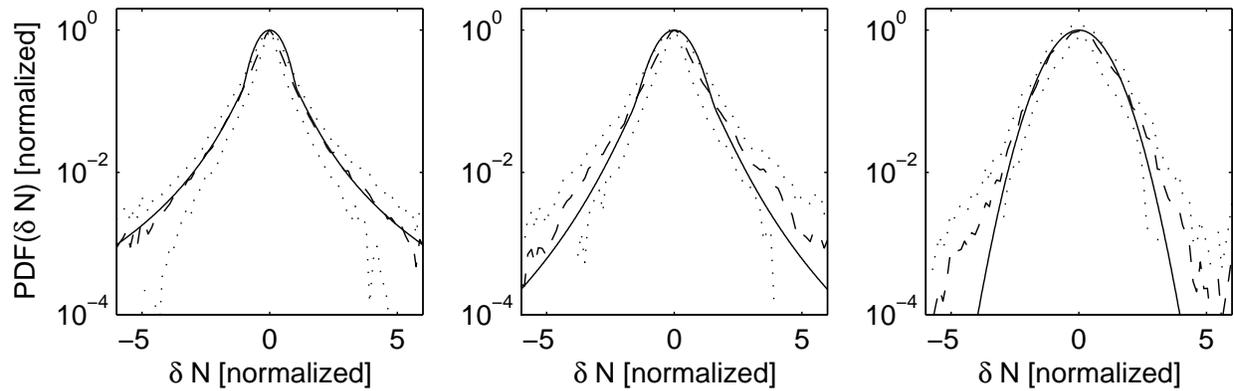}
\caption{Left panel: The PDF of the increments of observed WIND density
fluctuations for $\tau = 10$ and a resolution of 92 sec.
as compared to the bi-kappa function with $\kappa=2$. Based on the same
data the central panel provides the characteristics for increase $\tau = 70$
where $\kappa$ assumes a value of 3.5 for the best representation. The PDF
of large scale density fluctuations with $\tau=900$ are well modeled by a
Gaussian with $\kappa = \infty$, right panel.    
\label{fig5}}
\end{figure}

\clearpage 

\begin{figure}
\plotone{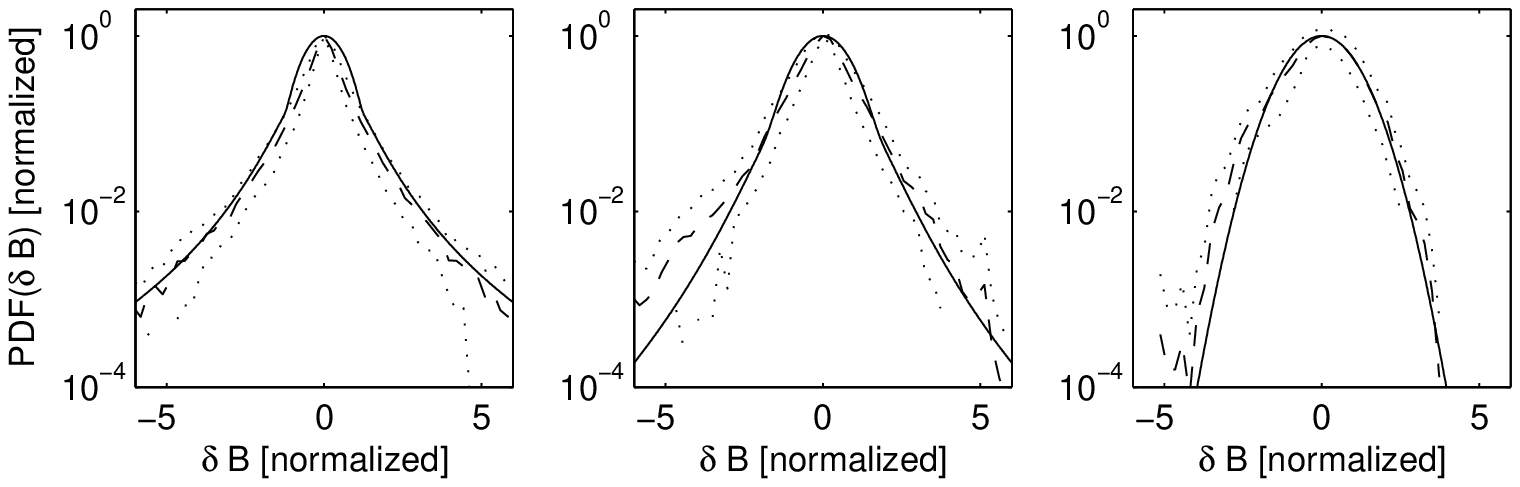}
\caption{Left panel: The PDF of the increments of observed WIND magnetic
field fluctuations for $\tau = 10$ and a resolution of 92 sec.
as compared to the bi-kappa function with $\kappa=2$. Based on the same
data the central panel provides the characteristics for increase $\tau = 70$
where $\kappa$ assumes a value of 3.5 for the best representation. The PDF
of large scale magnetic field fluctuations with $\tau=900$ are well modeled
by a Gaussian with $\kappa = \infty$, right panel.    
\label{fig6}}
\end{figure}

\clearpage 

\begin{figure}
\plotone{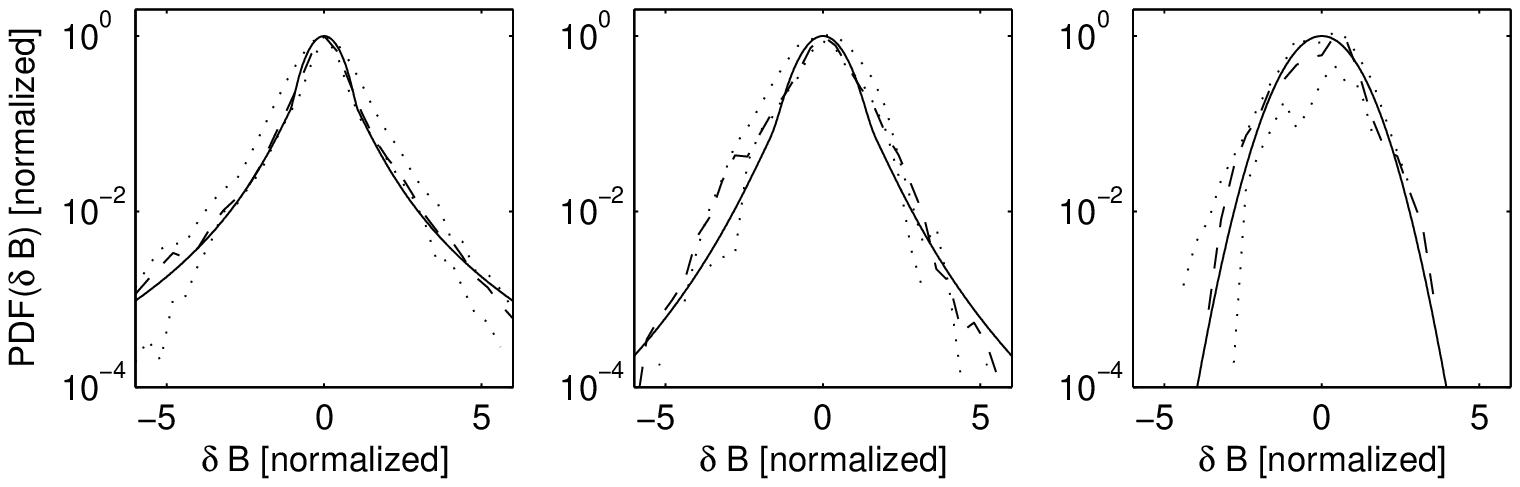}
\caption{Left Panel: the PDF of the increments of observed ACE magnetic
field fluctuations for $\tau = 100$ and a resolution of 16 sec.
as compared to the bi-kappa function with $\kappa=1.8$. Based on the same
data the central panel provides the characteristics for increased $\tau = 2000$
where $\kappa$ assumes a value of 3.0 for the best representation. The PDF of
large scale magnetic field fluctuations, $\tau = 10000$, are well modeled by
a Gaussian with $\kappa = \infty$, right panel.    
\label{fig7}}
\end{figure}

\end{document}